\documentclass[acsnano, twocolumn, superscriptaddress,showpacs]{revtex4-1}

\usepackage{amsmath}
\usepackage{amstext}
\usepackage{graphicx}
\usepackage{rotating}
\usepackage{tabularx}
\usepackage[below]{placeins}
\usepackage{float}
\usepackage{color}
\usepackage[english]{babel}
\usepackage{times}
\usepackage{xspace}

\def\avg#1{\left\langle#1\right\rangle}

\newcommand{\BS}{Bi$_2$Se$_3$\xspace}

\newcommand{\kv}{{\bf k}}
\newcommand{\w}{\omega}
\newcommand{\bc}{{\bf c}}

\begin{document}

\title{Anomalous Magneto Optic Effects from an Antiferromagnet Topological-Insulator Heterostructure}

\author{Amrit De}
\thanks{amritde@gmail.com, These authors contributed equally}
\author{Tonmoy K. Bhowmick}
\thanks{tbhow001@ucr.edu, These authors contributed equally}
\affiliation{Department of Electrical Engineering, University of California - Riverside, CA 92521}
\author{Roger K. Lake}
\thanks{rlake@ece.ucr.edu}
\affiliation{Department of Electrical Engineering, University of California - Riverside, CA 92521}

\date{\today}

\begin{abstract}
Materials with no net magnetization are generally not magneto-optically active.
While this is individually true for a collinear antiferromagnet (AFM) and a topological insulator (TI),
it is shown here that the magneto-optic Kerr effect (MOKE) emerges when the TI and AFM films are proximity coupled.
Because of the lack of macroscopic magnetization,
the AFM only couples to the spin of one of the TI's surfaces
breaking time-reversal and inversion symmetry -- which leads to a small $\mu$deg MOKE signal.
This small MOKE can be easily enhanced by 5 orders of magnitude, via cavity resonance,
by optimizing the AFM and TI film thicknesses on the substrate.
For slightly off-resonant structures, a $6^\circ$ Kerr rotation can be electrically switched on by varying the Fermi energy.
This requires less than 20 meV, which is encouraging for low power spintronics and magneto-optic devices.
We further show that this simple structure is easily resilient to 5 \% material growth error.
\end{abstract}

\maketitle
\date{\today}


\begin{figure}
\centering
\includegraphics[width=0.75\columnwidth]{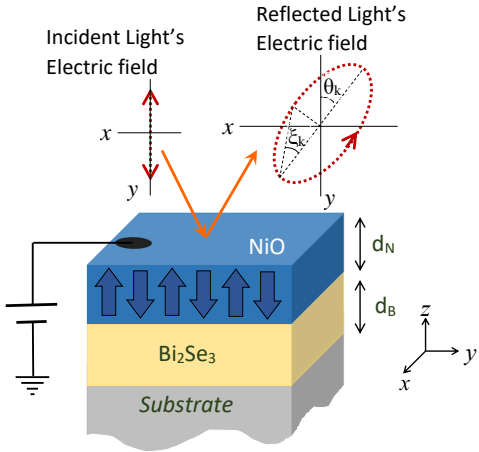}
\caption{MOKE from a thin-film collinear Antiferromagnet on top of a thin-film Topological-Insulator. We specifically consider a NiO film on Bi$_{2}$Se$_{3}$ thin-film grown on a SiO$_2$ substrate.
}
\label{fig:TiAfm}
\end{figure}

The Faraday effect and the magneto optic Kerr effect (MOKE) can be viewed as optical manifestations of the Berry curvature.
The optical response functions in MOKE are analogous to those of the electrical Hall conductivity.
However, there are subtle differences and additional probing capabilities due to the
inter- and intra-band transitions, interface effects, and frequency- and directional dependencies.
These properties lead to magneto-optic effects
in quantum Hall devices\cite{Kukushkin1996, ronen2018robust},
quantum materials such as topological insulators (TIs) \cite{Tse2010,Tse2011,serra2018observation},
magnetised chiral systems\cite{Sessoli2015},
and Skyrmions\cite{Vomir2016,Jiang2016,woo2018current,Jiang2017,Bhowmick2018}.
Magneto-optic effects are useful for device applications\cite{Arima2008,atmatzakis2018magneto}.
Optical isolators exploit the Faraday effect, and
polar MOKE is used to optically read out magnetically stored information\cite{Hansen1990,Suzuki1992,Mansuripur2000}.
Even with large macroscopic magnetizations, the Kerr rotation tends to be small -- barely one degree,
for most magneto-optic (MO) materials
\cite{nvemec2018antiferromagnetic,Lairson1993apl,Van1983apl,Egashira1974jap,Reim1988apl,duine2018synthetic}.
Some exceptions such as CeSb \cite{Pittini1996prl} rely on resonance effects from the hetrostructure.


MOKE requires broken time reversal symmetry (TRS), which is usually achieved by
the presence of an external magnetic field or macroscopic magnetization.
%
Strongly magnetized materials, such as ferromagnets (FMs) show large MOKE,
but they also have large stray magnetic fields that are undesirable for devices.
%
%
Antiferromagnets (AFMs) are more attractive since they have negligible stray fields and also electrically
switch much faster that FMs \cite{Wadley2016,Jungwirth2016,Baltz2018}.
However a collinear AFM with no stray magnetization generally will not show MOKE,
although there can be exceptions that we discuss below \cite{MOKE_Coll_AFM_DXiao_PRL16}.
Typically, a TI that preserves TRS will not display MOKE,
although a few $\mu$deg Faraday rotations from a TI's surface were recently
detected using nonlinear magneto-optics and circularly polarized pumps \cite{Mondal2018}.
Observing MOKE in TIs generally requires either external magnetic fields \cite{Tse2010,Tse2011},
magnetic doping \cite{Patankar2015},
or proximity coupling to materials with net magnetization \cite{Lang2014,Bhowmick2018}.
%

In this paper we show that a MOKE signature arises at the interface of a TI and a collinear g-type AFM.
%
%
This MOKE signature can be enhanced by 5 orders of magnitude, to 1-2$^\circ$,
by using resonant enhancements from the structure and small electric fields.
This strong anomalous MOKE occurs in the absence of external magnetic fields,
stray magnetic moments, or magnetic dopants.

The MOKE in this system is a result of the breaking of symmetries
that allow MOKE and the anomalous Hall effect (AHE) in non-collinear
AFMs with no net magnetic moments
\cite{MacDonald_AHE_NC-AFM_PRL14,
QNiu_MOKE_NC-AFMs_PRB15,
AHE_NC-AFM_Nat15,
Parkin_AHE_NC-AFMs_ScAdv16,
MOKE_NC-AFM_CLChien_NatPhot19,
Crooker_MOKE_NC-AFM_APL19,
MOKE_Mn3Ge_APL20},
and the predicted voltage controlled MOKE in a collinear AFM \cite{MOKE_Coll_AFM_DXiao_PRL16}.
In a three dimensional TI, TRS is preserved, and
the degenerate Kramers pairs of the surface states exist on opposing surfaces.
In a g-type AFM, TRS is broken by the opposing spin alignments on its bipartite lattice,
and, macroscopically, there is no net magnetization.
Proximity coupling the AFM to one surface of the TI, breaks both TRS and inversion symmetry,
and this allows the system to exhibit a non-zero MOKE with no net magnetic moment.


In this paper, we specifically consider a heterostrcuture as illustrated
in Fig. \ref{fig:TiAfm} with material parameters corresponding to
a NiO-film (AFM) grown on a Bi$_{2}$Se$_{3}$-film (TI) deposited on a SiO$_2$ substrate.
The resulting Kerr rotation from a TI single surface is tiny ($\sim\mathcal{O}$[$\mu$deg]) as expected.
However huge enhancements, resulting in 5 orders of magnitude increase in the MOKE,
are shown to arise by carefully choosing the film thicknesses of NiO and \BS.
The NiO layer and the SiO$_2$ substrate form a cavity, which can boost the MOKE via cavity resonance effects\cite{Pittini1996prl}. The resonant optical frequency is film thickness dependent.
We further show that Kerr rotations of over $5^\circ$ can be obtained by electrically biasing slightly detuned structures.
In general, expected film growth errors will naturally detune the MOKE resonance. We show that for $\pm$ 5 \% film-thickness errors, even in the worst case,
at least 1 deg Kerr-rotations can still be obtained by applying an electrical bias of under 20 meV.
%
Overall this simple and practical TI:AFm device can generate huge MOKE, while consuming very little power. It is planar, compact and free of stray magnetic fields and external magnets.
These features are very attractive for practical spintronic devices, sharp MO switches, MO memory and electro-optics including optical isolators.
%

\begin{figure}
\centering
\includegraphics[width=1\columnwidth]{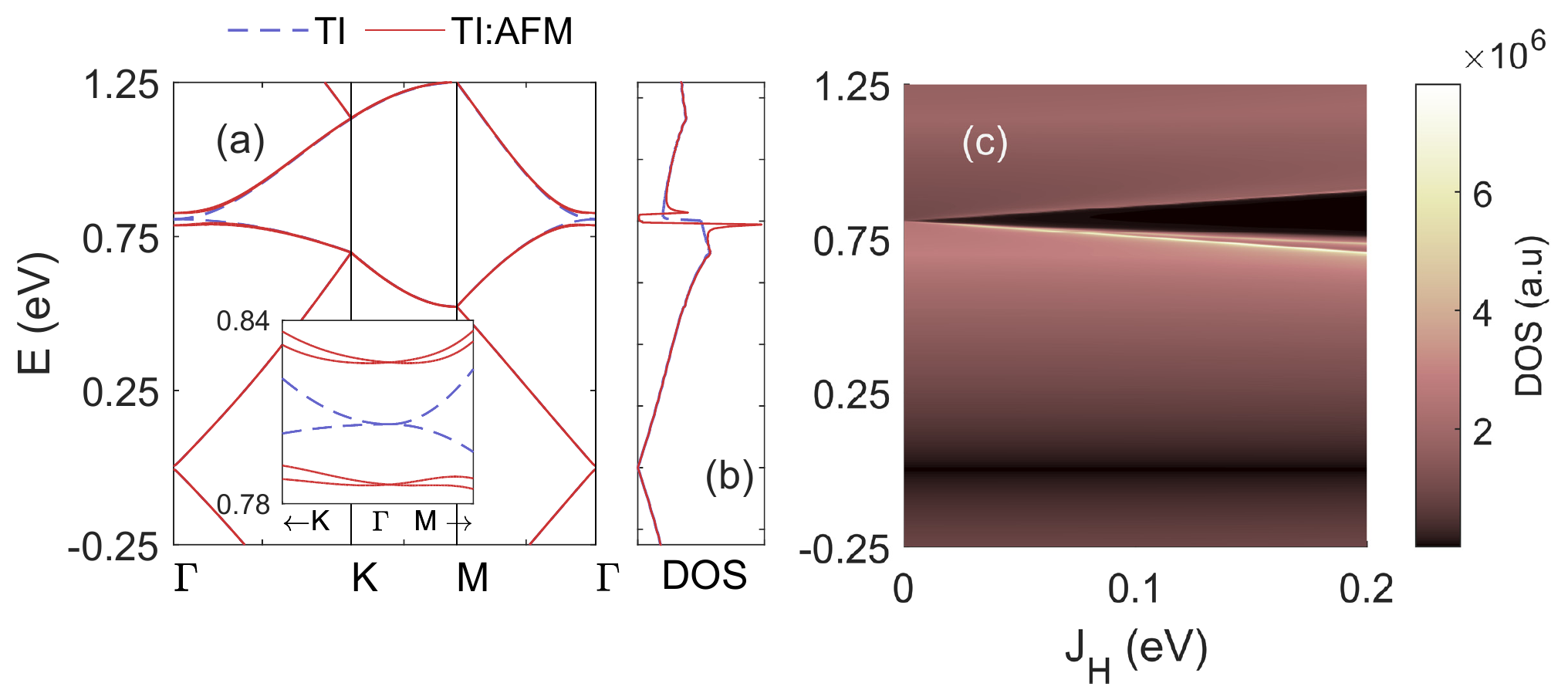}
\caption{{(a)} Bandstructure and {(b)} density of states (DOS) for TI only ($J_H = 0$, dashed blue line)
and for TI:AFM ($J_H = 40$ meV, solid red line).
The inset in (a) shows the higher energy gap and the Rashba type dispersion that occurs with $J_H = 40$ meV.
{(c)} DOS as a function of $J_H$.
}
\label{fig:E-k}
\end{figure}

\begin{figure}
  \includegraphics [width=1\columnwidth]{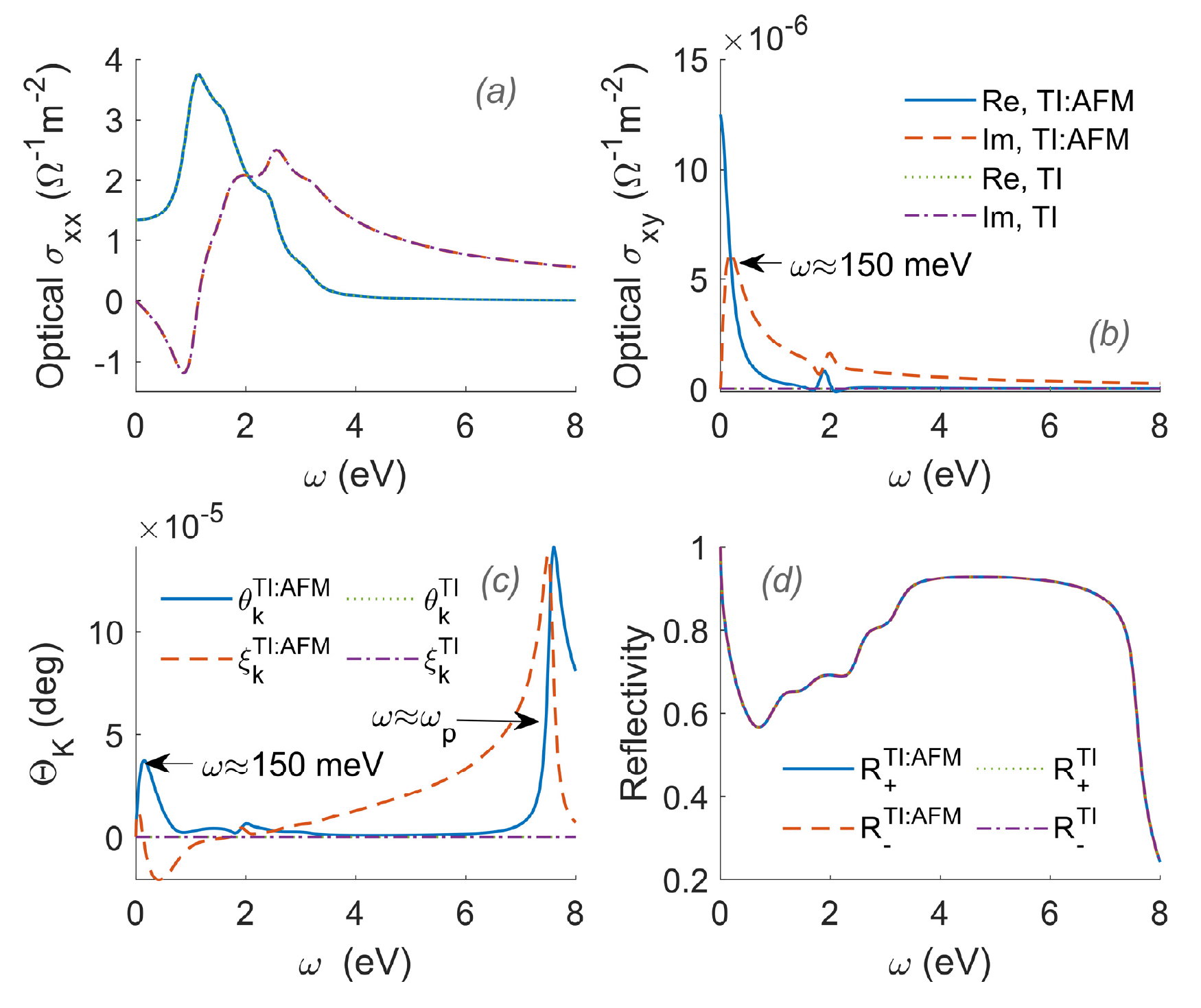}
  \caption{Real and imaginary parts of the {(a)} diagonal- and {(b)} off-diagonal optical conductivity for the TI alone
and a coupled TI / AFM. Note that (a) and (b) share the same legend.
{(c)} Kerr rotation and ellipticity and {(d)} reflectivities for TI alone and for a single AFM:TI interface
as denoted in the legend.}
   \label{fig:SI}
\end{figure}


\emph{The Model:}
The low energy zone-center effective Hamiltonian for a thin-film TI \cite{Liu2010,Zeng2018} is
$H_{0}(\boldsymbol{k}) = \tau_{z}h_{D}(\boldsymbol{k}) + m_{k} \tau_{x}$,
where $\tau_z$ and $\tau_x$ are Pauli matrices, respectively representing the TI's top and bottom surfaces and the hybridization between them.
$h_{D}(\boldsymbol{k}) =  \hbar v(k_{y}\sigma_{x} - k_{x}\sigma_{y})$ is the two-dimensional Dirac cone Hamiltonian with Fermi velocity $v$ and $m_{k} = m_{0} + m_{1} (k_{x}^2 + k_{y}^2)$ is the interlayer hybridization.
%
%
We discretize this Dirac model and obtain the following tight-binding Hamiltonian for an AFM proximity coupled to a TI,
\begin{equation}\label{eq:h2}
 H = \sum_{i}\bc_{i}^{\dagger} {h_{i}} \bc_{i} +
\sum_{\avg{i,j}} (\bc_{i}^{\dagger} {\bf t} \bc_{j} + h.c.) +
J_{H}\sum_{i}\bc_{i}^{\dagger} \boldsymbol{\sigma}'_{i} \cdot \boldsymbol{S}_{i} \bc_{i}.
\end{equation}
where ${{h}}_{i} = (m_{0} + \frac{4 m_{1}}{a^2}) \tau_x \otimes \sigma_0$,
the site indices $\avg{i,j}$ run over all nearest neighbor sites,
and
${\bf c}_i = [ c_{i,1,\uparrow} \; c_{i,1,\downarrow} \; c_{i,2,\uparrow} \; c_{i,2,\downarrow}]^T$
is the spinor
annihilation operator for site $i$ and layers $1$ and $2$.
Here $t\in\{t_x,t_y\}$ represents nearest neighbor hopping where $t_{x(y)} =   \pm\frac{i \hbar v}{2a}\tau_z\otimes\sigma_{y(x)} - \frac{m_{1}}{a^2}\tau_x\otimes I$.
It is implied that $\boldsymbol{\sigma}'=I\otimes{\boldsymbol\sigma}$,
where $I$ is the identity, $\boldsymbol\sigma=\{\sigma_x,\sigma_y,\sigma_z\}$ is the Pauli spin vector for the TI's itinerant electron and $\boldsymbol{S}=\{S_x,S_y,S_z\}$ is the spin-vector of the AFM.
The TI's surface spins interact with the AFM's spins via the Hund's rule coupling term $J_{H}$.



The electronic bandstructure (shown in Fig. \ref{fig:E-k}(a)) and wavefunctions for the
TI:AFM system are numerically calculated using a 2 $\times$ 2 supercell,
where the N\'{e}el vector of the AFM texture is perpendicular to the TI's surface.
The TI's top surface is proximity coupled to a G-type AFM thin-film.
We assume periodic boundary conditions along $x$ and $y$.
We set $m_{0} = 6$ meV, $m_{1} = 0.2$ eV$\text{\AA}^2$, and $v = 0.5 \times 10^{6}$ m/s.
The discretization length $a = 10 \text{\AA}$  and $J_H = 40$ meV.
The Dirac cone is trivially gapped at $\Gamma$ due to the $m(\kv)$ term,
and this gap is unaffected by the proximity coupling to the AFM.

The proximity coupling of the G-type AFM to the TI has several effects.
First, it increases the periodic unit cell from a single tight-binding site consisting of
two surfaces and 4 spins to 4 tight-binding sites with two surfaces and 16 spins
resulting in 16 bands.
This doubling of the unit cell causes zone folding of the Brillouin zone resulting in
crossing of bands at $\Gamma$ at higher energies.
The TI's proximity coupling to the AFM with $J_H = 40$ meV does not affect the low energy levels near
the Dirac point.
However a Rashba type gap opens at the higher band crossing, as shown in Fig. \ref{fig:E-k} (a) and (b),
and this energy gap increases linearly with $J_H$ as shown in Fig. \ref{fig:E-k}(c).
Since the bands are symmetric around $E=0$, a similar gap also opens below the Fermi level.
The gapping and Rashba type dispersion at the higher band crossing results in singularities
in the density of states on either side of the gap, as can be seen in Fig. \ref{fig:E-k}(b).
These bands, resulting from
broken inversion symmetry, time reversal symmetry, and zone-folding,
are then involved in the optical transitions
which lead to the magneto optic properties of this system.



\emph{Dielectric Tensor Components:}
Magneto optic effects are determined by the dielectric tensor which depends on the band-structure and its topology.
In particular for the polar Kerr effect considered here,
the N\'{e}el vector of the AFM is along $z$, which is perpendicular to the surface and parallel
to the optical incidence.
The $x$ and $y$ directions preserve in-plane symmetry.  The complex $3\times 3$ dielectric tensor has
$[\epsilon_{xx},\epsilon_{yy},\epsilon_{zz}]$ diagonal terms and off-diagonal $\epsilon_{xy}$ terms which are topology dependent.

The matrix elements of the optical conductivity tensor are obtained from the Kubo formula
\cite{Callaway_74,Ebert_MO_review96},
\begin{widetext}
\begin{eqnarray}{
\sigma_{\mu\nu} =  \frac{ie^2}{h L} \displaystyle\int \frac{d^2 k}{(2\pi)^2} \sum_{n,l}
\frac{ f_{nl}(\boldsymbol{k}) }{\omega_{nl}(\boldsymbol{k})}
\left(
\frac{\Pi(\boldsymbol{k})_{nl}^{\mu}\Pi(\boldsymbol{k})_{ln}^{\nu} }{ \omega - \omega_{nl}(\boldsymbol{k}) + i\gamma }
+
\frac{  \Pi(\boldsymbol{k})_{ln}^{\mu}\Pi(\boldsymbol{k})_{nl}^{\nu} }{ \omega + \omega_{nl}(\boldsymbol{k}) + i\gamma }\right),
\label{eq:Kubo}}
\end{eqnarray}
\end{widetext}
where $\Pi^{\mu}_{nl}(\boldsymbol{k}) = \langle\psi_n(\boldsymbol{k})|v_{\mu}|\psi_l(\boldsymbol{k})\rangle$ is the
matrix element of the velocity operator, $v_\mu = \frac{\partial H}{\hbar \partial k_\mu}$, where
$\{\mu,\nu\}\in\{x,y\}$.
The energy broadening parameter $\gamma = 17.5$ meV for all calculations.
$\hbar \omega_{nl}(\boldsymbol{k}) = E_{n}(\boldsymbol{k}) - E_{l}(\boldsymbol{k})$ is the energy
difference of an optical transition between an unoccupied band $n$ and an occupied band $l$.
$f_{nl}(\boldsymbol{k})=f_n(\boldsymbol{k})-f_l(\boldsymbol{k})$ where $f_n(\boldsymbol{k})$ is the Fermi factor,
however all calculations are performed at zero temperature.
$L$ is the thickness associated with the TI surface state taken to be 1 nm.

The velocity operator is similar to the Berry connection.
For a single surface this leads to a momentum space gauge potential or an equivalent magnetic field.
For a TI, the gauge fields for each surface cancel each other unless the symmetry between
the top and bottom surfaces is broken, such as by an AFM on one side.
The resulting MO effects can therefore be viewed as an optical manifestation of the Berry curvature.


Since an effective Hamiltonian has been used to obtain $\sigma_{\mu\nu}$,
the missing higher band contributions 
are compensated for by adding a $\kappa / (\omega+i\gamma)$ term
to the optical dielectric tensor as follows:
$\epsilon_{\mu\nu}(\omega) = \varepsilon_o\delta_{\mu\nu} - \frac{4\pi i}{\omega}\sigma_{\mu\nu} -\frac{\kappa}{\omega + i\gamma}$,
where $\varepsilon_o$ is the vacuum permittivity.
$\kappa$ is adjusted so that the relative zero frequency dielectric constant $\epsilon_0$
(obtained from the optical sum rules) matches the known experimental value
\cite{Eddrief2016} for Bi$_2$Se$_3$.
%


For calculating polar-MOKE, the required complex in-plane refractive index is ${n}_\pm = \sqrt{\epsilon_\pm}=\sqrt{\epsilon_{xx}\pm i\epsilon_{xy}}$ where, the $+(-)$ signs
represents right(left)
circularly polarized (RCP(LCP)) light propagation.
The complex MOKE is,
$\Theta_k = \mathcal{\theta}_k +i\xi_k$
where the Kerr rotation and ellipticity respectively are,
\begin{eqnarray}
\theta_k &=& (\Delta_+-\Delta_-)/2 \\
\xi_k &=& (|r_+|-|r_-|)/(|r_+|+|r_-|) .
\end{eqnarray}
Since the eigen-modes here are LCP and RCP, the Kerr rotation angle can be expressed as the phase difference between these two modes.
The complex phase $\Delta_{\pm}$ is in turn obtained from the Fresnel reflection coefficients $r^\pm = |r^\pm|\exp(-i\Delta_\pm)$.
The observed reflective intensity is $R_\pm=|r^\pm|^2$.
The MOKE in a multilayer thin-film structure can be significantly altered by internal reflection at various interfaces.
At normal incidence, $r^\pm$ for $N$ thin-films can be calculated using a $2\times2$ characteristic matrix method \cite{Bhowmick2018,De2002a} in the LCP/RCP eigenmode basis:
\begin{eqnarray}\label{TM}
\boldsymbol{S}^\pm =
\prod_{j=0}^{N}
\frac{1}{t_{j,j+1}^\pm}
  \left[
\begin{array}{cc}
  1 &  r_{j,j+1}^{\pm} \\
    r_{j,j+1}^{\pm} &  1 \\
\end{array}
\right]
  \left[
\begin{array}{cc}
  e^{i\beta_{j+1}^\pm} & 0 \\
  0 &  e^{-i\beta_{j+1}^\pm} \\
\end{array}
\right]
\end{eqnarray}
where $\beta_j = (2\pi/\lambda)n_j d_j$ is a phase factor, $d_j$ is the thickness of the $j^{th}$ layer and $\lambda$ is the optical wavelength.
The Fresnel reflection and
transmission coefficients at normal incidence for each interface respectively are
$r_{j,j+1}^\pm = (n_j^\pm-n_{j+1}^\pm)/(n_{j}^\pm + n_{j+1}^\pm)$ and
$t_{j,j+1}^\pm = (2 n_j^\pm)/(n_{j}^\pm + n_{j+1}^\pm)$.
The resultant complex reflection coefficient is
$r^\pm = S_{12}^\pm/S_{11}^\pm = |r^\pm|\exp(-i\Delta_\pm)$ where $S_{\mu\nu}^\pm\in\boldsymbol{S}^\pm$.
%

\begin{figure}
\includegraphics [width=1\columnwidth]{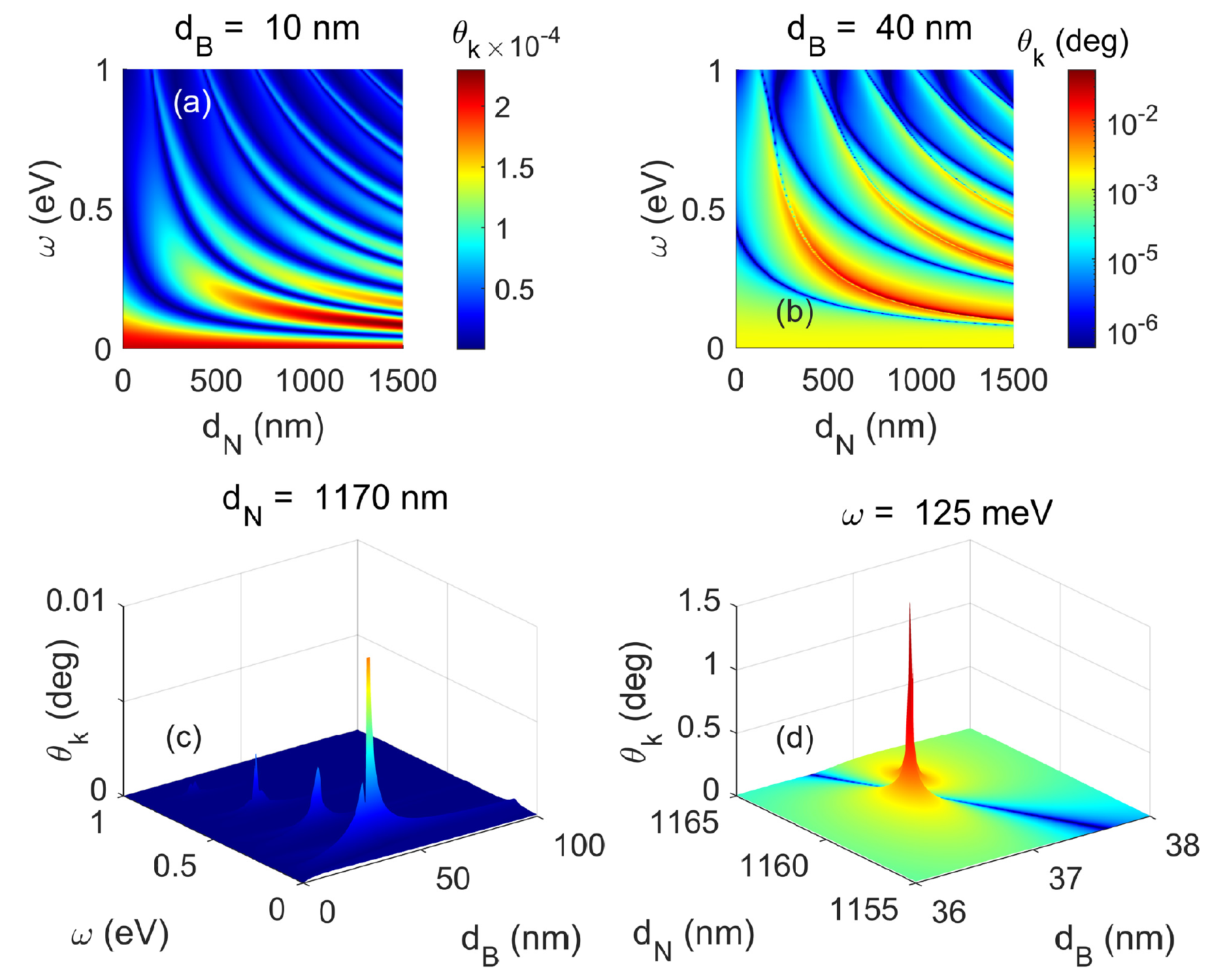}
\caption{
The Kerr-rotation ($\theta_k$) phase diagram shown as a function of
(a) $d_{N}$ and $\omega$ for $d_{B} = 10$ nm,
(b) $d_{N}$ and $\omega$ for $d_{B} = 40$ nm,
(c) $d_{N}$ and $\omega$ for $d_{B} = 1.17 \mu m$
and (d) $d_{N}$ and $d_{B}$ at $\omega = 125$ meV.
}
\label{fig:Phase1}
\end{figure}


\emph{Discussion:}
The calculated optical dielectric functions 
of the TI with- and without the AFM layer are shown in Fig. \ref{fig:SI}(a,b).
For the TI alone, the numerically calculated $\sigma_{xy}$ is zero ($\sim \mathcal{O} (10^{-64})$ -- well below numerical precession).
Once an AFM is introduced on one side of the TI,
the resulting $\sigma_{xy}$ increases to
$\mathcal{O} (10^{-6})~\Omega^{-1}m^{-2}$ as shown in Fig. \ref{fig:SI}(b).
However there is no change in the diagonal optical conductivity, $\sigma_{xx}$, as shown in Fig. \ref{fig:SI}(a).
This is a significant result even though $\sigma_{xy} \ll \sigma_{xx}$.
%

%
%

These effects can be directly observed using MOKE.
The resulting Kerr rotations, elipticity, and reflectivities for the TI:AFM
are shown in Fig. \ref{fig:SI}(c,d) for a single interface.
The Kerr rotation features can be understood by examining the approximate expression for complex MOKE:
$\Theta_k  \approx {\epsilon_{xy}}/\left(\sqrt{\epsilon_{xx}}(1-\epsilon_{xx})\right)$ (since $\epsilon_{xy} \ll \epsilon_{xx}$) \cite{Argyres1955}.
The MOKE resonance with $\theta_k\approx 4\times 10^{-5}$ degrees occurs at
$\omega(\theta_k^{max})\approx$ 150 meV in the low energy regime as shown in
Fig. \ref{fig:SI}(c), which directly corresponds to the $\sigma_{xy}$ peak in Fig. \ref{fig:SI}(b).
There is also a high energy MOKE resonance that occurs at 7.5 eV.
Using the $1-\epsilon_{xx}$ resonance condition and the Drude model it can be shown that this is near the plasma frequency $\omega_p$ \cite{Feil1987,De2002,De2003,Abe2004,Bhowmick2018}.
We extracted $\omega_p$=7.32 eV from $\sigma_{xy}$ using the optical sum rules \cite{Bhowmick2018}.
Similarly, our extracted cyclotron frequency $\omega_c$, from $\sigma_{xy}$, is 3.15 $\mu$eV, which is the effective $\omega_c$ of a single TI surface.



MOKE can be enhanced by the resonance effects that arise from optimizing the film thickness of different materials.
In order to understand the effects of this for our system,
we consider a thin-film structure as shown in Fig. \ref{fig:TiAfm}, where a NiO
film of thickness $d_N$ sits on a Bi$_{2}$Se$_{3}$ film of thickness $d_B$.
The transfer matrices, Eq. (\ref{TM}), are used to calculate the MOKE spectra of the multi-layer structures, assuming
normal incidence and in-plane material isotropy.
The optical dispersion relations for NiO were obtained from literature \cite{Sriram2013}.
The AFM:TI effects manifest themselves via $n_\pm$.
The refractive index of air is used for the semi-infinite media above the NiO layer,
and the dispersive refractive indices of SiO$_2$ were used
for the semi-infinite substrate \cite{Palik.Handbook}.

For the TI:AFM device, the Kerr rotation angle's phase diagram is shown in Fig. \ref{fig:Phase1} as a function of $\omega$, $d_B$ and $d_N$.
The complete phase space is quite large.
Therefore, representative phase diagrams are shown in Fig. \ref{fig:Phase1} where one of the parameters is held constant and the other two are varied.
In Fig. \ref{fig:Phase1}(a) and (b), $d_B$ is held at 10 nm and 40 nm respectively.
The Kerr rotation angle $\theta_k$ has a considerable dependence on the NiO film thickness as it creates the resonances in the TI:AFM structure. For $d_B =~40$ nm, $\theta_k$ reaches 0.5 deg for $d_N\approx$ 1200 nm.
%
%
As shown in Fig. \ref{fig:Phase1}(c), with the NiO thickness fixed at its near optimal value,
a number of $\theta_k$ resonances begin to appear, with the highest peak being at \BS thicknesses of about 40 nm at $\w$ = 125 meV.
Note that the $\theta_k$ resonances occur at harmonics of $\omega$, which is mainly dictated by $d_N$.
Finally, with $\w$ fixed at $125$ meV, the $\theta_k$ phase diagram is shown in Fig.
\ref{fig:Phase1}(d) as a function of $d_N$ and $d_B$.
An notable and sharp $\theta_k$ resonance happens at $d_N=$ 1162 nm and $d_B=$ 37 nm.
The maximum Kerr rotation reaches 1.5 deg (for 1 {\AA} thickness resolution) --
an enhancement of 6 orders of magnitude compared to the value from the single surface.
The ability to tune $\theta_k$ to exceed 1 deg, is quite remarkable given that this is a purely anomalous MOKE that manifests in a collinear TI:AFM structure with
no net magnetization and no external magnetic field.

\begin{figure}
\includegraphics [width=1\columnwidth]{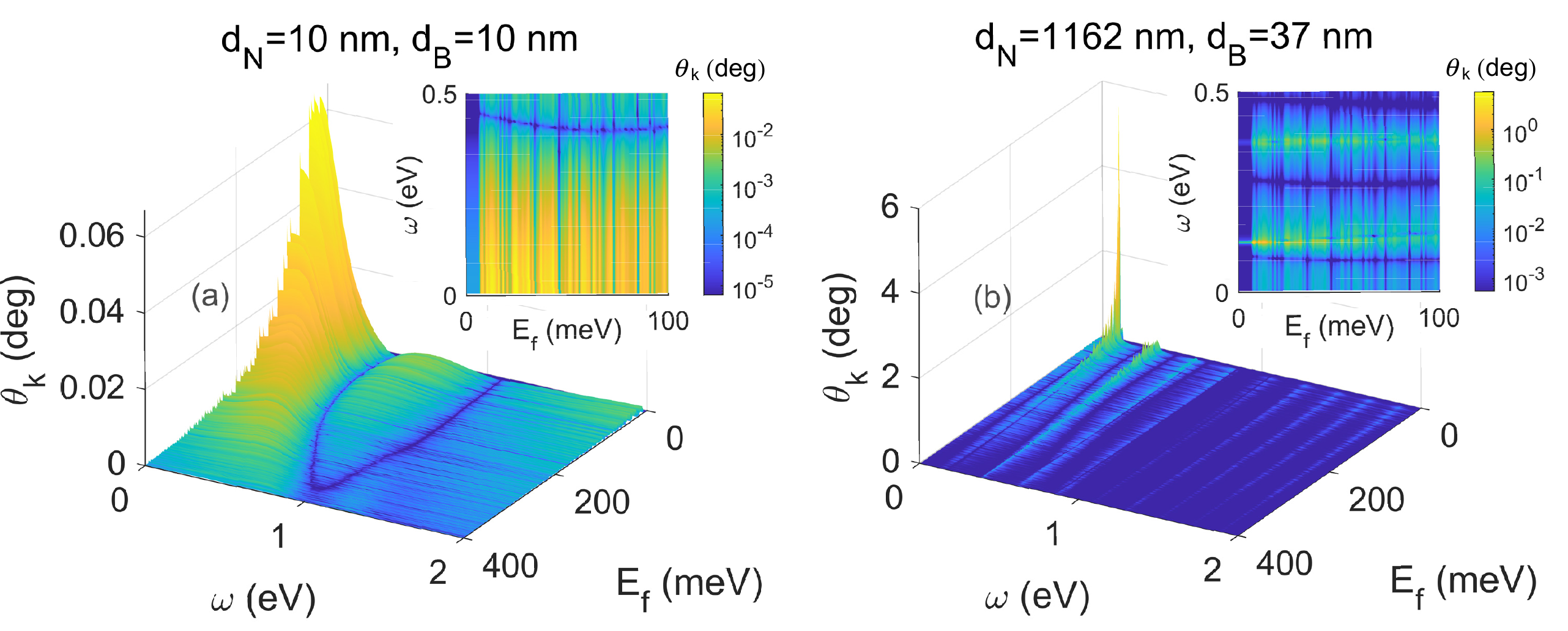}
\caption{The Kerr rotation angle as a function of Fermi energy, $E_f$, and optical frequency, $\omega$ for {\bf(a)} $d_B = 10$ nm and $d_N=10$ nm and {\bf(b)} $d_B=1.16~\mu$m and $d_N=100$ nm. The insets are zoomed-in to highlight the sudden switching behavior as a function of $E_f$. }
\label{fig:Phase2}
\end{figure}

Next, we analyze the Fermi level dependent MOKE spectra for the TI:AFM structure
in Fig. \ref{fig:Phase2}.
This phase diagram is particularly important for electro-optic device applications
as it provides an electrical handle to control the anomalous magneto-optic effect in the TI:AFM.
$\theta_k$ is shown as a function of Fermi energy, $E_f$ and $\w$ for two cases: $\{d_N,d_B\}=\{10,10\}$ nm and $\{1162,37\}$ nm.
Although, the \{10,10\}-nm case is not optimized, it experimentally easier to grow,
it is easier to electrically gate, and its physics is easier to explain even
though the MOKE is quite weak.
The second optimized thicker structure will have added cavity
induced resonance effects but with very strong MOKE.
In Fig. \ref{fig:Phase2}(a), there is a sudden jump in $\theta_k$ when $E_f$ crosses the bandgap.
As the Fermi level is further swept through the conduction bands's Dirac cone, the number of optical transitions steadily decrease, which makes $\theta_k^{max}$ shift to higher frequencies.
These physical trends are further superposed on top of the strong cavity
induced resonances as shown in Fig. \ref{fig:Phase2}(b) for the $\{1162,37\}$ nm structure.
The MOKE resonances will only appear at optical harmonics as determined by the
cavity (formed by NiO and the substrate).
At $E_f=$7 meV, $\theta_k\sim$ 6 deg resonance occurs.
This $\theta_k$ resonance is still over 4 deg by $E_f=$8 meV.

\begin{figure}
\includegraphics [width=1\columnwidth]{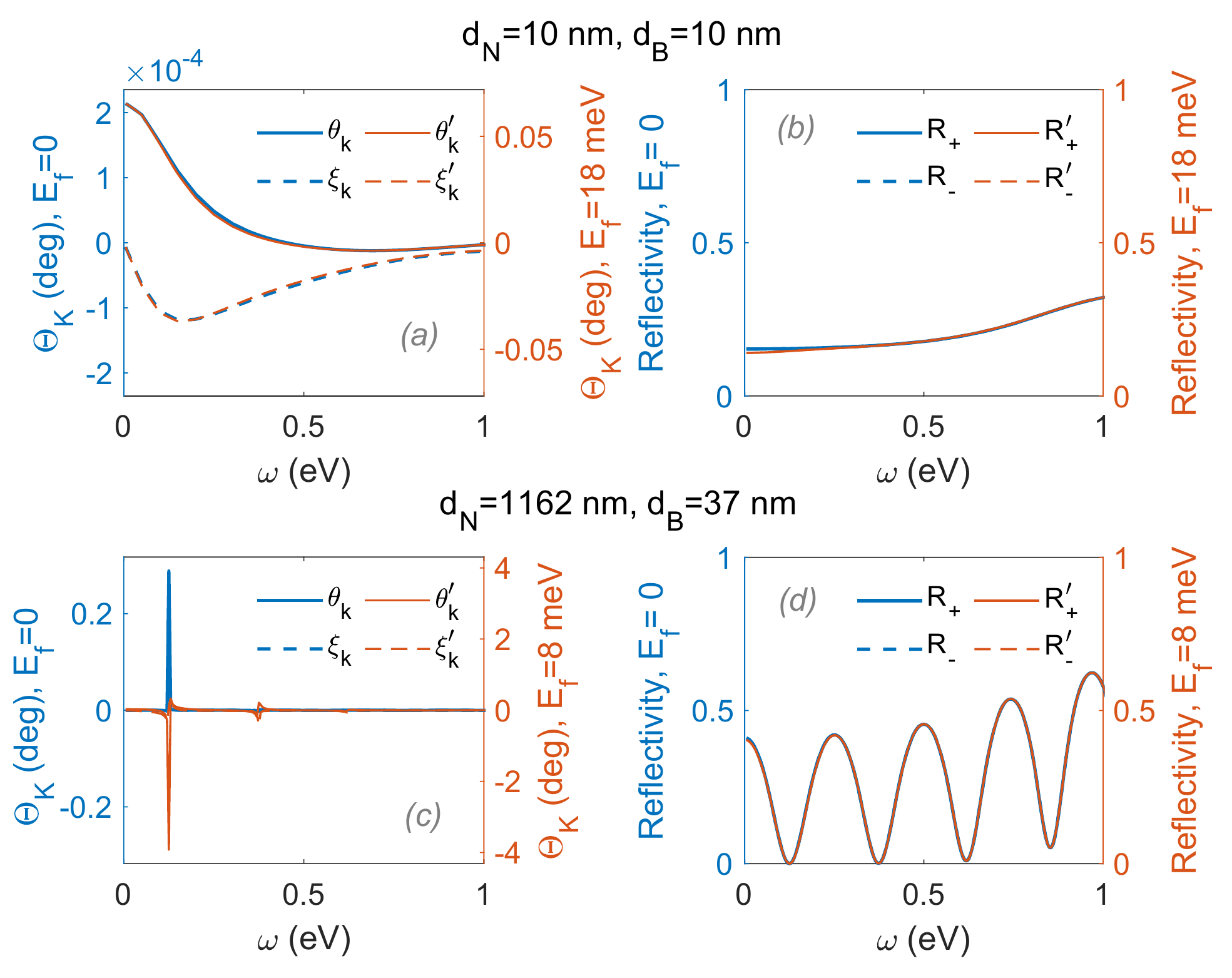}
\caption{We show the {\bf(a)} Kerr rotation and ellipticity and {\bf(b)} the reflectivity spectra for $d_N=d_B=10$ nm. Next we show the {\bf(c)} MOKE spectra and {\bf(d)} reflectivity spectra for $d_N=1162$ nm and $d_B=37$ nm. The left (right) y-axis on all the plots is for $E_f=0$ ($E_f\neq 0$).
}
\label{fig:TF}
\end{figure}
\begin{figure}
\includegraphics [width=1\columnwidth]{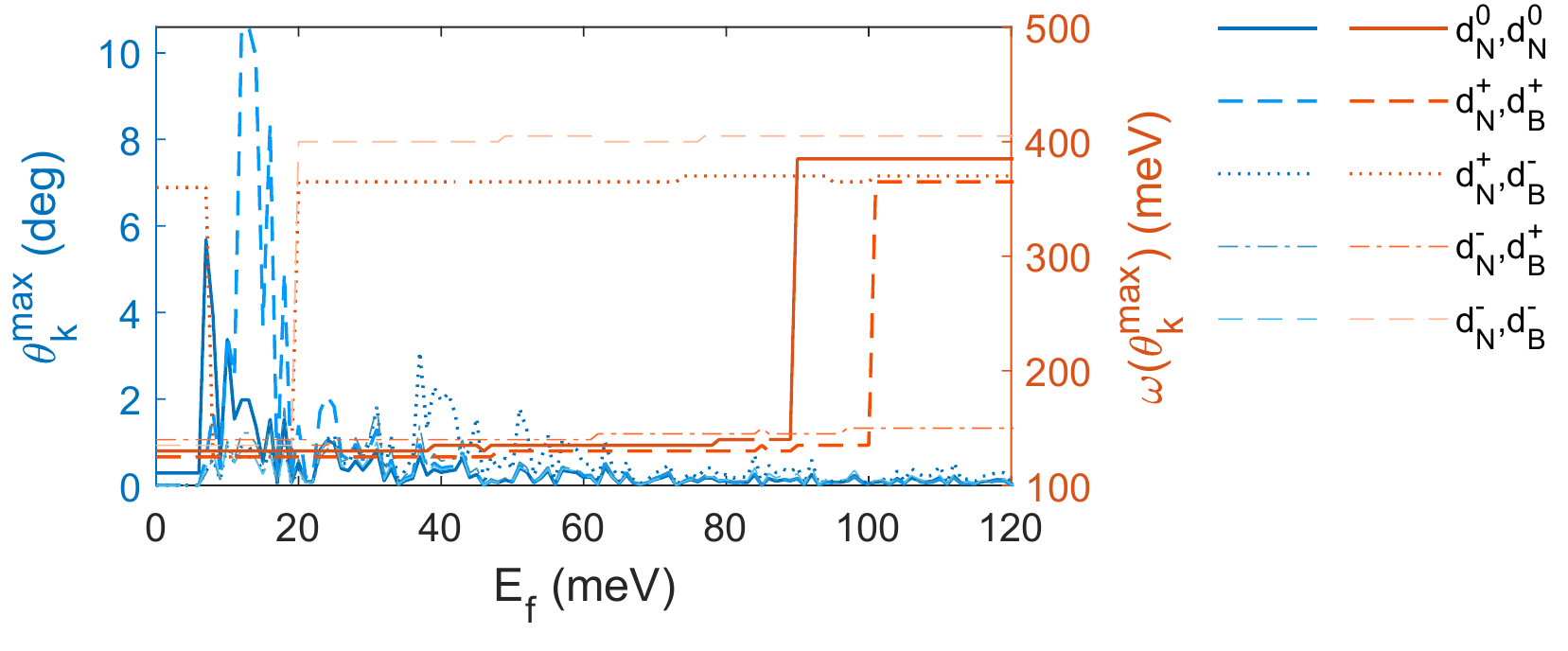}
\caption{
{(left-axis)} The maximum Kerr rotation ($\theta_k^{max}$) as a function of Fermi energy ($E_f$) for structures with $\pm$ 5\% errors in layer thicknesses.
Here $\{d_N^{0},d_B^{0}\}$=\{1162,37\} nm and $d_{N(B)}^\pm = (1\pm 0.05)\times d_{N(B)}^0$.
{(right-axis)} The $\w$ at which $\theta_k^{max}$ occurs as a function of $E_f$.
}
\label{fig:Hall}
\end{figure}

We further analyze the MOKE spectra and the reflectivity for the two cases
($\{d_N,d_B\}=\{10,10\}$ nm and $\{1162,37\}$ nm) in Fig. \ref{fig:TF}.
%
%
In addition we also examine the MOKE as a function of two different Fermi energies ($E_f$).
The left (right) y-axis on all plots in Fig. \ref{fig:TF} is for $E_f$=0 (optimal bias).
Our choice of the optimal bias for each structure was based on the results of Fig. \ref{fig:Hall}.
As compared to the MOKE from the single TI:AFM interface shown in Fig. \ref{fig:SI},
the low-frequency peak in the Kerr rotation spectra is enhanced by 4 to 5 orders of magnitude by the resonance effects that arise from the thin-film thicknesses and interfaces.
A number of subsequent smaller MOKE resonances appear at higher harmonics of the fundamental peak at $\omega = 125$ meV as expected.
Maximum $\theta_k$ is also accompanied by a corresponding dip in the reflectivity.
%
For the thinner-films in Fig. \ref{fig:TF}(a), the MOKE decreases monotonically as a function of $\omega$, as expected for the valance- and conduction-band states on the TI's Dirac-cone.
This monotonic behavior is absent for the thicker optimized $\{1162,37\}$ nm TI:AFM structure as shown in Fig. \ref{fig:TF}(c) and (d).
Instead, multiple MOKE spectral resonances now appear at harmonics of $\w$ that are resonant with the optical-cavity like structure.
The maximum Kerr rotation is around $0.3$ deg in Fig. \ref{fig:TF}(c), for $E_f$=0, after rounding off $\{d_N,d_B\}$ to the nearest nm.
The frequency at which the maximum Kerr rotation occurs, $\w(\theta_k^{max})$, mainly shifts with $d_N$ as shown in Fig.\ref{fig:Phase1}-(b). These resonant like enhancements can still be observed for \emph{growth errors} in layer-thicknesses.



The Kerr angle $\theta_k$ can be sharply boosted by 1 to 3 orders of magnitude
by applying a small bias to raise the Fermi energy in the TI:AFM,
as shown in Fig. \ref{fig:TF} (right-axis).
For $E_F$=18 meV, $\theta_k$ exceeds 0.05 deg for the nm-regime thin-films.
Remarkably, $\theta_k$ reaches $4^\circ$ for the thicker optimized structure with a small 8 meV bias.
Ignoring the sign difference, the MOKE spectra also scales up uniformly
in this small-bias regime with the $E_f$=0 results in Fig. \ref{fig:TF}.
The related reflectivity changes in \ref{fig:TF} (b) and (c) are quite small and not notable. 


Finally, since thin-film growth can be prone to errors during the material deposition process, it is important to analyse how the TI;AFM device would behave with notable errors in the film thicknesses.
In Fig. \ref{fig:Hall}, the maximum Kerr rotation angle ($\theta_k^{max}$) is shown as a function of Fermi energy with $\pm$ 5 \% thickness-errors.
Here $d_{N(B)}^\pm$ denotes $\pm$ 5\% deviation from the optimized $d_{N(B)}$.
The right-axis in Fig. \ref{fig:Hall} shows the $\w$ at which $\theta_k^{max}$ occurs.
Nominally, the Kerr rotation is nearly 6 deg at an 7 meV bias for the error free case.
While in the best case, a $+$ 5\% error in the layer-thicknesses leads to an enhancement of the MOKE with $\theta_k^{max}\approx$ 10 deg with a 12 to 14 meV bias.
For the worst case, $\theta_k^{max}$ is about 1 deg with an 8 meV bias.
This is very encouraging since a 1 deg Kerr rotation is still quite large. This shows that even with errors in layer growth near the resonance condition, it is still quite possible to observe a substantial anomalous MOKE signal by just electrically tuning the TI:AFM.
In addition this is the basis of a very useful electro-optic device.

A notable feature is that $\w(\theta_k^{max})$ increases in a distinct step-like manner.
The step size depends on the film-thicknesses and can be explained from Fig. \ref{fig:Phase2}.
The MOKE resonances only occur at higher harmonics of $\w_r$ (where $\w_r \equiv \w(\theta_k^{max})$ at $E_f=0$).
As the Fermi level sweeps the Dirac-cone conduction band,  $\theta_k^{max}$ tends to shift to higher $\w$, due to steadily declining optical transitions.
However, because of the thin-film structure, $\theta_k^{max}$ can only jump to a higher harmonic of $\w_r$ as shown in Fig. \ref{fig:Phase2}(b),
which leads to the $\w(\theta_k^{max})$ steps in Fig. \ref{fig:Hall}.

In summary, the MOKE arising from a thin film NiO / \BS heterolayer structure
displays very interesting physical properties with potentially important device applications.
Experimentally measurable Kerr rotations arise in the AFM-TI structure even though neither the AFM nor the TI have any net magnetization.
The collinear AFM's proximity to one of the TI surfaces leads to the breaking
of both time reversal symmetry and inversion symmetry.
This results in a small but observable MOKE signature.
The polar-MOKE geometry is best suited for observing this effect,
since the light is incident on the AFM-TI interface.
This small MOKE can be enhanced 5 orders of magnitude
by optimizing the AFM and TI film thicknesses,
which leads to a cavity resonance condition where the AFM and the substrate form a natural cavity.
For slightly off-resonant structures,
a 6 deg Kerr rotation can be obtained by varying the Fermi energy.
This is encouraging for practical low power devices as the Fermi energy
has to be varied by less than 20 meV.
We further show that this simple structure is resilient to 5 \% material growth error.
Overall this can lead to practical low-power spintronics, fast electro-optic switches, magneto-optic memory and gate controlled optical-isolators.
This device is simple to grow, there is no magnetic doping required, it is planar, compact and free of stray magnetic magnetization and external magnets.

\emph{Acknowledgement:}
This work was supported as part of Spins and Heat in Nanoscale Electronic Systems (SHINES) an
Energy Frontier Research Center funded by the U.S. Department of Energy, Office of Science,
Basic Energy Sciences under Award \#DE-SC0012670.
%

\end{document}